\documentclass[nonblindrev]{informs3} 

\usepackage{balance}
\usepackage{amsmath}
\usepackage[tight,footnotesize]{subfigure}
\usepackage{graphicx}
\usepackage{bm}
\usepackage{algorithm}
\usepackage[noend]{algorithmic}
\usepackage{multirow}
\usepackage{slashbox}
\usepackage{caption}
\usepackage{amsfonts}
\usepackage{siunitx}

\OneAndAHalfSpacedXII 

\usepackage{natbib}
 \bibpunct[, ]{(}{)}{,}{a}{}{,}%
 \def\bibfont{\small}%
\TheoremsNumberedThrough     

\EquationsNumberedThrough    

\MANUSCRIPTNO{2015} 

\begin{document}


\RUNAUTHOR{Yuan and Tang}

\RUNTITLE{Submodular Participatory Budgeting}

\TITLE{Submodular Participatory Budgeting}

\ARTICLEAUTHORS{%
\AUTHOR{Jing Yuan}
\AFF{Department of Computer Science and Engineering, University of North Texas}
\AUTHOR{Shaojie Tang}
\AFF{Naveen Jindal School of Management, The University of Texas at Dallas}
} 

\ABSTRACT{Participatory budgeting refers to the practice of allocating public resources by collecting and aggregating individual preferences. Most existing studies in this field often assume an additive utility function, where each individual holds a private utility for each candidate project, and the total utility of a set of funded projects is simply the sum of the utilities of all projects. We argue that this assumption does not always hold in reality. For example, building two playgrounds in the same neighborhood does not necessarily lead to twice the utility of building a single playground.

To address this, we extend the existing study by proposing a submodular participatory budgeting problem, assuming that the utility function of each individual is a monotone and submodular function over funded projects. We propose and examine three preference elicitation methods, including \emph{ranking-by-marginal-values}, \emph{ranking-by-values} and \emph{threshold approval votes}, and analyze their performances in terms of distortion. Notably, if the utility function is addicative, our aggregation rule designed for threshold approval votes achieves a better distortion than the state-of-the-art approach.}


\maketitle

\section{Introduction}
Participatory budgeting \citep{cabannes2004participatory}, which allocates public resources by collecting and aggregating individual preferences subject to a budget constraint, has attracted increased attention recently. It has been shown that participatory budgeting has successfully allocated over \$170 million of public funds to more than 500 local projects, primarily in the US, Canada and Europe \citep{benade2021preference}. At a high level, this problem belongs to a bigger categolray of a central societal question on how to aggregate individuals' preferences to make a collecative decision \citep{brandt2016handbook}. The input of such a problem consists of  a set of alternatives and a group of voters, each voter is allowed to provide its preference over those alternatives- typically through ranking, the goal of this problem is to select a set of winning alternatives that best reflect all individuals' preferences \citep{caragiannis2017subset}.  While most of existing studies in this field overlooks the cost associated with each alterative, \citep{lu2011budgeted} extends this study to a setting called budgeted social choice. In their setting, each alterantive has a cost and the selected alterantives must satisfy a budget constraint.

Building upon the framework established by \citep{goel2019knapsack,benade2021preference}, we can decompose any participatory budgeting approach into two components. The first component is the \emph{input format}, which refers to how each voter's preferences are elicited. The second component is the \emph{aggregation rule}, which pertains to how these preferences are combined to select a group of alternatives. Some popular input formats include the \emph{ranking-by-values method}, where each voter ranks all alternatives according to their values, and the \emph{approval-based method}, where each voter submits a list of acceptable alternatives. These preferences can be aggregated using variants of classic voting rules, such as the Borda count rule.

\subsection{Our Model and Contributions}

Most existing studies in this field, including \citep{goel2019knapsack,benade2021preference}, assume an additive utility function, where each individual holds a private utility for each alternative, and the total utility of a set of alternatives is simply the sum of the utilities of these alternatives. We argue that this assumption  may not always hold true in practice, particularly when project interactions are present \citep{jain2020participatory}. For example, building two playgrounds in the same neighborhood does not necessarily lead to twice the utility of building a single playground.

To address this, we extend the existing study by proposing a submodular participatory budgeting problem, where we assume that each voter's utility function is monotone and submodular over the selected alternatives. This effectively captures the negative synergy among alternatives, as illustrated in the playground example above. Since every additive function is inherently submodular, we are tackling a more general problem in this paper.

We explore three preference elicitation methods: \emph{ranking-by-marginal-values}, \emph{ranking-by-values}, and \emph{threshold approval votes}, and evaluate their performance through the lens of implicit utilitarian voting \citep{caragiannis2017subset}. We adopt the notation of \emph{distortion} \citep{procaccia2006distortion,mandal2020optimal} to quantify the effectiveness of an aggregation rule in maximizing social welfare. Informally, the distortion is defined as the worst-case ratio of the social welfare achieved by an optimal outcome and the social welfare achieved by a given preference elicitation method.

For the ranking based methods, prior to our work, it was unclear how to rank projects when the utility is non-additive. One of our contributions is the development of a novel \emph{ranking-by-marginal-values} mechanism applicable to non-additive utility functions. We demonstrate that this method, as well as the classical \emph{ranking-by-values} method, achieves a distortion of
$O(\log m\sqrt{m}/(1-c))$ where $m$ is the number of alternatives and $c\in[0,1]$ is the curvature  of each submodular utility function. Intuitively, a larger value of $c$ indicates that a submodular function is closer to an additive function. Notably, for the case of addictive utility functions (i.e., $c=0$), our approaches achieve the same distortion (i.e., $O(\log m\sqrt{m})$) as the current state-of-the-art approach \citep{benade2021preference}.

Our another contribution is the development of an aggregation rule for \emph{threshold approval votes}. We show that this approach achieves a distortion of $O(\log m/(1-c))$. Notably, for additive utility functions (i.e., $c=0$), our approach achieves a better distortion (i.e., $O(\log m)$) than the current state-of-the-art result $O(\log^2 m)$.

\subsection{Additional Related Work}
Participatory budgeting has been an active field in computational social science \citep{aziz2017proportionally,talmon2019framework}. The most relevant work to ours is \citep{benade2021preference}, which examines the distortion of various preference elicitation methods, assuming that each voter's utility function is additive. Our study builds on and extends their work by considering a more general submodular setting.

While most existing studies in this field assume an additive setting, a few recent papers address \emph{project interactions} \citep{durand2024detecting}. For example, \citep{jain2020participatory} propose a model that captures the synergies between projects (a.k.a. alternatives), which are assumed to be known, through a partition over the projects. Specifically, projects within the same partition set either have a complementary effect (a positive interaction) or a substitution effect (a negative interaction). Our study differs from theirs in two key aspects. First, they assume that each voter's utility function is explicitly given. For instance, they define a voter's utility with respect to a set of funded alternatives as a function of the number of alternatives that are both approved by the voter and selected to be funded. In contrast, our setting falls under implicit utilitarian voting, meaning we do not assume access to each voter's utility function. Instead, voter preferences are expressed through their voting. This introduces additional challenges and necessitates the adoption of the notion of distortion to evaluate the performance of a participatory budgeting method. Second, they assume that the interaction structure is revealed through partitioning over the projects, and this information (i.e., partitioning) is publicly available. However, we do not assume such knowledge. Instead, we assume a very general submodular function for each voter, and most importantly, this information is private and known only to the voter. We note that our submodular setting can only capture the substitution effect (negative interaction) among alternatives. Investigating more general settings that also account for positive interactions is left as future work.

Another work closely related to ours is \citep{caragiannis2017subset}. Their goal is to select a set of \(k\) alternatives, where a voter's utility for a subset of alternatives is defined as the maximum utility for any alternative in the subset. It is easy to verify that their study is a special case of ours. This is because their size constraint is a specific instance of the knapsack constraint, and their utility function is a particular example of a monotone submodular function.
\section{Preliminaries: Submodular Function and Curvature}
 Given a function $f$, define the marginal utility of an item $i \in A$ with respect to a set of items $S \subseteq A$ as $f(i \mid S) \stackrel{\text{def}}{=} f(S \cup \{i\}) - f(S)$.

A function $f$ is said to be submodular if and only if for any two sets $X$ and $Y$ such that $X \subseteq Y$ and any item $i \notin Y$, $f(i \mid X) \geq f(i \mid Y)$. Furthermore, a submodular function $f$ is said to have \emph{curvature} $c \in [0,1]$ if $f(i \mid S) \geq (1 - c) f(\{i\})$ for any $S \subseteq A$ and $i \notin S$ \citep{balkanski2016power}. Curvature intuitively measures how much a given function deviates from a modular function.

\section{The Model}
 For any integer $k$, let $[k] = \{1, 2, \cdots, k\}$. The input of our problem is a set of $n$ \emph{voters} $[n]$, and a set $A$ of  $m$ \emph{alternatives} (e.g., projects). Each alternative $a\in A$ has a cost of $c_a$, and the total budget is $1$ (after normalization). For any subset of alternatives $S\subseteq A$, we abuse notation and let $c(S)=\sum_{a\in S} c_a$ denote the total cost of $S$. Define $\mathcal{F}=\{S\subseteq A: c(S)\leq 1\}$ as a collection of all feasible sets of alternatives.

We assume that each voter $i$ holds a utility function $f_i: 2^A \rightarrow \mathbb{R}_{\geq0}$, here we assume that each utility function $f_i$ is monotone and submodular. To ensure fairness among all voters, we follow the standard setting \citep{benade2021preference,boutilier2012optimal} and assume that $f_i(A) = 1$ for all $i\in [n]$. Given the \emph{utility profile} $\hat{f}=\{f_1, f_2, \cdots, f_n\}$, the (utilitarian) social welfare of a set of alternatives $S\subseteq A$ is defined as $\textsf{sw}(S, \hat{f})=\sum_{i\in[n]} f_i(S)$.

The utility function $f_i$ of a voter $i$ is not directly accessible, however, it can be accessed through $i$'s vote $\rho_i$ that is induced by $f_i$. Let $\hat{\rho} = \{\rho_1, \rho_2, \cdots, \rho_n\}$ denote the \emph{input profile}.
Let $\hat{f} \triangleright \hat{\rho}$ denote that the utility profile $\hat{f}$
is consistent with the input profile $\hat{\rho}$.
In this paper, we examine three  preference elicitation methods: \emph{group based ranking-by-marginal-values}, \emph{group based ranking-by-values} and \emph{threshold approval votes}, and analyze their performance through the lens of implicit utilitarian voting.

\subsection{Group based Ranking-by-Marginal-Values}
The group based ranking-by-marginal-values method asks each voter to rank \emph{partial} alternatives according to a simple greedy algorithm. This method aims to approximate the voter's true utility function by assuming that the order in which alternatives are added reflects their relative importance according to marginal utility.

First, building on the idea from \citep{benade2021preference}, we partition all alternatives into \(\log m + 1\) groups based on their costs. Specifically, let $l_0=0$ and $u_0=1/m$. For $t\in[\log m]$ (for ease of presentation, assume $m$ is a power of $2$), define $l_t=2^{t-1}/m$ and $u_t=2^t/m$. We partition all alternatives into $\log m+1$ groups based on their costs: $G_0=\{a\mid c_a\leq u_0\}$ and $G_t=\{a\in A \mid c_a\in(l_t, u_t]\}$ for $t\in [\log m]$.   We choose a group $G_t$ uniformly at random, and asks each voter to rank all alternatives in $G_t$ according to a simple greedy algorithm. This algorithm starts with an empty set $S=\emptyset$. Then for each alternative not yet ranked  $a \in G_t \setminus S$, it computes the marginal value $f_i(a \mid S)$, and adds $a_j = \arg\max_{a \in G_t \setminus S} f_i(a \mid S)$ to the ranking list. Here $a_j$ is simply the best alternative that has the highest marginal utility when added to $S$. The final ranking $\rho_i = (a_1, a_2, \ldots, a_{|G_t|})$ reflects the order in which alternatives were selected.

While the idea of partitioning was adopted from \citep{benade2021preference}, they use this technique solely for analysis; during their preference elicitation stage, voters are required to rank \emph{all} alternatives. In contrast, our method requires voters to rank only the alternatives from a randomly selected group $G_t$. This design has two advantages: first, it helps to lower the communication cost incurred during the elicitation process compared to the approach that ranks all alternatives; second, and most importantly, it achieves better distortion.

\subsection{Group based Ranking-by-Values}
Similar to the group based ranking-by-marginal-values method, our second approach, termed as group based ranking-by-values method, asks each voter to rank partial alternatives. The key difference in this method is that voters rank alternatives based on their stand-alone values.

Specifically, we follow the same procedure as described in the previous method to partition all alternatives into $\log m +1$ groups: $G_0, G_1, \cdots, G_{|\log m|}$. We choose a group $G_t$ uniformly at random, and asks each voter to rank all alternatives in $G_t$ according to their stand-alone values. That is, in the final ranking $\rho_i$  from voter $i\in[n]$, for any two alternatives $a\in G_t$ and $b\in G_t$, $a$ is placed ahead of $b$ if and only if $f_i(\{a\}) \geq f_i(\{b\})$.

This method is similar to the classical ranking method designed for additive utility functions \citep{benade2021preference,boutilier2012optimal}, but it does not require each voter to rank \emph{all} alternatives. This reduces the communication cost during the preference elicitation process. However, unlike the ranking-by-marginal-values method, collecting partial rankings does not offer the additional benefit of reducing distortion. This is because ranking all alternatives in their stand-alone values provides more information than ranking only a subset.

\subsection{Threshold Approval Votes} The \emph{threshold approval votes} method asks each voter to report all alternatives whose stand-alone utility is above a certain threshold $\alpha$.  Formally, given a threshold $\alpha$, the approval set for voter $i$ is given by:
\[ \rho_i = \{a \in A \mid f_i(\{a\}) \ge \alpha\}. \]

This method captures the notion that voters are willing to endorse (approve) alternatives that meet a minimum acceptable utility level, rather than providing a complete ranking. The term threshold approval votes was first introduced in \citep{benade2021preference} for additive functions. Our contribution is to develop a novel aggregation rule and use additional techniques to deal with submodular functions. Notably, for the special case of additive functions, our method achieves lower distortion  compared to the previous study.

\subsection{Problem Formulation} Following the implicit utilitarian voting framework, our goal is to maximize the (utilitarian) social welfare. We introduce the notation of \emph{distortion} \citep{procaccia2006distortion} to capture the distance between the social welfare achieved by an aggregation rule $\pi$ and the maximum social welfare. Formally, the distortion of $\pi$ with respect to a vote profile $\hat{\rho}$ is given by
\begin{eqnarray}
\textsf{dist}(\pi, \hat{\rho})= \sup_{\hat{f}: \hat{f} \triangleright \hat{\rho}}\frac{\max_{T\in \mathcal{F}} \textsf{sw}(T, \hat{f})}{\mathbb{E}[\textsf{sw}(\pi(\hat{\rho}), \hat{f})]}
\end{eqnarray}
where $\pi(\hat{\rho})$ denotes the set of alternatives selected by $\pi$. In this paper, we focus on a randomized participatory budgeting approach; hence, we are interested in the expected social welfare achieved by an aggregation rule $\pi$.

The (overall) distortion of a rule $\pi$ is defined as $\textsf{dist}(\pi) = \max_{\hat{\rho}} \textsf{dist}(\pi,\hat{\rho})$. Let $\pi^*$ denote the optimal (random) aggregation rule that minimizes the distortion. I.e.,
\begin{eqnarray}
\pi^* = \arg\min_{\pi} \textsf{dist}(\pi).
\end{eqnarray}

Note that distortion is defined with respect to a given preference elicitation method; hence, different methods might yield different minimum distortions. By thoroughly analyzing the distortion achieved under various preference elicitation methods, one can assess the potential of each method in accurately recovering individuals' actual preferences \citep{benade2021preference}.
\section{Distortion Analysis}
\subsection{Group based Ranking-by-Marginal-Values} We first recall the procedure of this approach. Let $l_0=0$ and $u_0=1/m$. For $t\in[\log m]$ (for ease of presentation, assume $m$ is a power of $2$), define $l_t=2^{t-1}/m$ and $u_t=2^t/m$. We partition all alternatives into $\log m+1$ groups based on their costs: $G_0=\{a\in A\mid c_a\leq u_0\}$ and $G_t=\{a\in A \mid c_a\in(l_t, u_t]\}$ for $t\in [\log m]$.

We choose a group $G_t$ uniformly at random, and asks each voter to rank all alternatives in $G_t$ according to a simple greedy algorithm. This algorithm starts with an empty set $S=\emptyset$. Then for each alternative not yet ranked  $a \in G_t \setminus S$, it computes the marginal value $f_i(a \mid S)$, and adds $a_j = \arg\max_{a \in G_t \setminus S} f_i(a \mid S)$ to the ranking list. Here $a_j$ is simply the best alternative that has the highest marginal utility when added to $S$. The final ranking $\rho_i = (a_1, a_2, \ldots, a_{|G_t|})$ reflects the order in which alternatives were selected.

We follow the general framework of \citep{benade2021preference} and introduce some additional notations. Consider a ranking $\rho$ and an alternative $a \in G_t$. We define $\rho(a)$ as the position of $a$ in $\rho$. For an input profile $\hat{\rho}$, the \emph{harmonic score} of $a$ in $\hat{\rho}$ is given by $\textsf{sc}(a, \hat{\rho}) = \sum_{i=1}^n 1/\rho_i(a)$. 

Let $\hat{f}$ denote the \emph{true} utility profile that is consistent with $\hat{\rho}$ and let $S^*=\arg\max_{S\in \mathcal{F}} \textsf{sw}(S, \hat{f})$ denote the optimal set of alternatives that maximizes the social welfare.

Next, we describe our aggregation rule, which randomly picks a rule from the following two candidates. This aggregation rule was first introduced in \citep{benade2021preference} to address the case of additive functions; however, additional efforts are required in our analysis to tackle the challenges posed by non-additive functions.

\begin{itemize}
\item Rule A: Let $G^+_t$ denotes the top (up to) $\sqrt{m}\cdot(1/u_t)$ alternatives from $G_t$ that have the largest harmonic scores in $\hat{\rho}$. Select a subset of $G^+_t$ with a size of $1/u_t$, also chosen uniformly at random.
\item Rule B: Choose a single alternative uniformly at random.
\end{itemize}

For the sake of analysis, we introduce some additional notations. Let $G^-_t=G_t\setminus G^+_t$. Note that because $\sqrt{m}\cdot(1/u_0)=\sqrt{m}\cdot m \geq m$, we have $G^+_0=G_0$. Define $G^+=\cup_{t=0}^{\log m} G^+_t$ and $G^-=A\setminus G^+$. We next present two lemmas to bound the social welfare achieved by Rule A and Rule B respectively.
\begin{lemma}
\label{lem:y}
The expected social welfare achieved by Rule A is at least  $\textsf{sw}(G^+ \cap S^*, \hat{f})/(1+\log m)\sqrt{m}$.
\end{lemma}
\emph{Proof:} Observe that Rule A chooses each group $G_t$ with probability $1/(1+\log m)$, then it selects each alternative in $G^+_t$ with probability at least $1/\sqrt{m}$ (this is because $|G^+_t|\leq \sqrt{m}\cdot(1/u_t)$ and Rule A selects a subset of $G^+_t$ with a size of $1/u_t$). Therefore, Rule A selects each alternative in $G^+$ with a probability of at least $1/(1+\log m)\sqrt{m}$.

We next present a result that links random sampling to submodular maximization.
\begin{lemma}
\label{lem:2}
(Lemma 2.2 of \citep{feige2011maximizing}). Consider a monotone submodular set function $g : 2^U \rightarrow \mathbb{R}_{\geq 0}$. Let $U(p)$ denote a subset of $U$ where each element is included with a probability of at least $p$ (elements are not necessarily independent of each other). The expected value of $g(U(p))$ is at least $p$ times the value of $g(U)$. Formally, $\mathbb{E}[g(U(p))] \geq p \cdot g(U)$.
\end{lemma}

Recall that $\textsf{sw}(S, \hat{f})=\sum_{i \in [n]} f_i(S)$ and $f_i$ is a monotone submodular function for each $i \in [n]$. Hence, $\textsf{sw}(\cdot, \hat{f})$ is also a monotone submodular function by the fact that a linear combination of monotone submodular functions is still a monotone submodular function. Recall that Rule A selects each alternative in $G^+$, with probability at least $1/(1+\log m)\sqrt{m}$. By substituting $g(\cdot)$ with $\textsf{sw}(\cdot, \hat{f})$ and substituting $p$ with $1/(1+\log m)\sqrt{m}$ in Lemma \ref{lem:2}, we have that the returned solution from Rule A has social welfare at least $\textsf{sw}(G^+, \hat{f})/(1+\log m)\sqrt{m}$, which is no less than $\textsf{sw}(G^+ \cap S^*, \hat{f})/(1+\log m)\sqrt{m}$ by the assumption that  $\textsf{sw}(\cdot, \hat{f})$ is a monotone function. $\Box$

We next show that the expected social welfare achieved under Rule B is at least $(1-c)\textsf{sw}(G^- \cap S^*, \hat{f})/2(1+\log m)\sqrt{m}$. While our proof follows the general framework of Theorem 3 in \citep{benade2021preference}, we must make additional efforts to address the challenges posed by non-additive functions, as previously mentioned.
\begin{lemma}
\label{lem:x}
Assume the  curvature of each function $f_i$ is at least $c$, the expected social welfare achieved by Rule B is at least $(1-c)\textsf{sw}(G^- \cap S^*, \hat{f})/2(1+\log m)\sqrt{m}$.
\end{lemma}
\emph{Proof:}   Observe that $G^-_0=\emptyset$. Therefore,
\begin{eqnarray}
\label{eq:17}
\textsf{sw}(G^- \cap S^*, \hat{f}) = \textsf{sw}(\cup_{t\in[\log m]}(G^-_t \cap S^*), \hat{f}).
\end{eqnarray}

Fix $t\in [\log m]$ and $a\in G^-_t$. Let $H_k$ be the $k$-th harmonic number, it is easy to verify that
\begin{eqnarray}
\sum_{b\in G_t} \textsf{sc}(b, \hat{\rho}) = n\cdot H_{|G_t|} \leq n\cdot H_{m}.
\end{eqnarray}
Recall that $G^+_t$ contains  the top  $\sqrt{m}\cdot(1/u_t)$ alternatives from $G_t$ that have the largest harmonic scores in $\hat{\rho}$, we have
\begin{eqnarray}
\label{eq:09}
 \textsf{sc}(a, \hat{\rho})  \leq \frac{n\cdot H_{m}}{\sqrt{m}\cdot(1/u_t)} < \frac{n\cdot (1+\log m)}{\sqrt{m}\cdot m/2^t}.
\end{eqnarray}

Meanwhile, for each voter $i\in[n]$ and alternative $b\in G_t$, we have
\begin{eqnarray}
\label{eq:1}
\Delta_i (b) \leq 1/\rho_i(b)
\end{eqnarray}
where $\Delta_i (b) = f_i(T \cup \{b\}) - f_i(T)$ (here $T=\cup_{a\in G_t: \rho_i(a) \leq \rho_i(b)} a$ denotes $b$'s proceeders in $\rho_i$) denotes the marginal utility of $b$ over all its proceeders in $\rho_i$. This is because $f_i(G_t)=\sum_{b\in G_t} \Delta_i (b, G_t) \leq 1$ and $\Delta_i (b, G_t)$ is no larger than the marginal utility brought by any of $b$'s predecessors in $\rho_i$ (this is by the submodularity of $f_i$ and the design of Ranking-by-Marginal-Values).

It follows that for each alternative  $a\in G^-_t$,
\begin{eqnarray}
\sum_{i\in[n]}\Delta_i (a, G_t) \leq \sum_{i\in[n]} 1/\rho_i(a)= \textsf{sc}(a, \hat{\rho}) < \frac{n\cdot (1+\log m)}{\sqrt{m}\cdot m/2^t}
\end{eqnarray}
where the first inequality is by inequality (\ref{eq:1}) and the second inequality is by inequality (\ref{eq:09}). This, together with the assumption that the  curvature of each function $f_i$ is at least $c$, implies that
\begin{eqnarray}
\label{eq:12}
(1-c)\sum_{i\in[n]} f_i(\{a\}) \leq \sum_{i\in[n]}\Delta_i (a, G_t) < n\cdot (1+\log m)/(\sqrt{m}\cdot m/2^t).
 \end{eqnarray}

 Moreover, by the definition of $G_t$, we have that each alternative  $a\in G^-_t$ has a  cost of at least  $l_t=2^{t-1}/m$. Therefore, for any $t\in[\log m]$, the ``value-to-cost'' density of each alternative $a\in G^-_t$ is at most
\begin{eqnarray}
&&\frac{\textsf{sw}(a, \hat{f})}{l_t} = \frac{\sum_{i\in[n]} f_i(\{a\})}{2^{t-1}/m}\leq   \frac{\frac{n\cdot (1+\log m)}{(1-c)\sqrt{m}\cdot m/2^t}}{2^{t-1}/m} = \frac{2n\cdot (1+\log m)}{(1-c)\sqrt{m}}
\end{eqnarray}
where the inequality is by inequality (\ref{eq:12}).

Notice that the above inequality applies to all alternatives in $\cup_{t\in[\log m]}(G^-_t \cap S^*)$, hence,
\begin{eqnarray}
&&\sum_{a\in \cup_{t\in[\log m]}(G^-_t \cap S^*)}  \textsf{sw}(a, \hat{f}) \leq \frac{2n\cdot (1+\log m)}{(1-c)\sqrt{m}}.
\end{eqnarray}
This is because the ``value-to-cost'' density of each alternative in  $G^-$ is at most $\frac{(1-c)2n\cdot (1+\log m)}{\sqrt{m}}$ and the total cost of $\cup_{t\in[\log m]}(G^-_t \cap S^*)$ is at most one (by the observation that $S^*$ is a budget feasible solution). This, together with inequality (\ref{eq:17}), implies that
\begin{eqnarray}
&&\textsf{sw}(G^- \cap S^*, \hat{f})= \textsf{sw}(\cup_{t\in[\log m]}(G^-_t \cap S^*), \hat{f}) \\
&&\leq \sum_{a\in \cup_{t\in[\log m]}(G^-_t \cap S^*)}\textsf{sw}(a, \hat{f})\\
&&\leq \frac{2n\cdot (1+\log m)}{(1-c)\sqrt{m}} \label{eq:117}
\end{eqnarray}
where the first inequality is by the assumption that $\textsf{sw}(\cdot, \hat{f})$ is a submodular function.

We are now ready to prove this lemma. Recall that according to Rule B, a single alternative is chosen uniformly at random. Therefore, each alternative has a probability of $1/m$ of being selected.  This, together with Lemma \ref{lem:2} and the assumption that $\textsf{sw}(\cdot, \hat{f})$ is a monotone submodular function, implies that
the expected social welfare achieved by  Rule B is at least $\textsf{sw}(A, \hat{f})/m=n/m$. Combining this with inequality (\ref{eq:117}), we have that the expected social welfare achieved by Rule B is at least $(1-c)\textsf{sw}(G^- \cap S^*, \hat{f})/2(1+\log m)\sqrt{m}$. $\Box$

Now we are in position to present the main theorem of this section.
\begin{theorem}
\label{thm:1}
Assume the  curvature of each function $f_i$ is at least $c$, the expected social welfare achieved by our aggregation rule is at least $(1-c)\textsf{sw}(S^*, \hat{f})/4(1+\log m)\sqrt{m}$. This also implies that the distortion of our approach is at most $O(\log m\sqrt{m}/(1-c))$.
\end{theorem}
\emph{Proof:} Recall that our aggregation rule randomly picks a rule from Rule A and Rule B at uniform, moreover,  the expected social welfare of Rule A is at least $\textsf{sw}(G^+ \cap S^*, \hat{f})/(1+\log m)\sqrt{m}$ (by Lemma \ref{lem:y}) and the expected social welfare of Rule B is at least $(1-c)\textsf{sw}(G^- \cap S^*, \hat{f})/2(1+\log m)\sqrt{m}$ (by Lemma \ref{lem:x}). Hence, the expected social welfare of our aggregation rule is at least
\begin{eqnarray}
&&\frac{\textsf{sw}(G^+ \cap S^*, \hat{f})/(1+\log m)\sqrt{m}+(1-c)\textsf{sw}(G^- \cap S^*, \hat{f})/2(1+\log m)\sqrt{m}}{2}\\
&&\geq (1-c)\frac{\textsf{sw}(G^+ \cap S^*, \hat{f})+\textsf{sw}(G^- \cap S^*, \hat{f})}{4(1+\log m)\sqrt{m}}\geq\frac{ (1-c)\textsf{sw}(S^*, \hat{f})}{4(1+\log m)\sqrt{m}}
\end{eqnarray}
where the second inequality is by the observations that $G^+ \cup G^- = A$ and $\textsf{sw}(\cdot, \hat{f})$ is a submodular function. $\Box$

\paragraph{Remark 1} Note that selecting a rule from Rule A and Rule B uniformly may not be optimal; one can certainly optimize this selection probability to reduce the distortion.

\paragraph{Remark 2} Interestingly, if we require each voter to rank all alternatives instead of only the alternatives from a random group, it might increase the distortion. This can be partially explained by the submodularity of the utility function. For example, ranking all alternatives based on their marginal values makes it harder to disentangle the correlations among alternatives, thereby making it more difficult to assess the ``true'' value of each individual alternative. This contrasts with our observation in the next method (ranking by stand-alone values), where providing a global ranking is always beneficial.

\subsection{Group based Ranking-by-Values} Recall that under this method, we follow the same procedure as described in the previous method to partition all alternatives into $\log m +1$ groups: $G_0, G_1, \cdots, G_{|\log m|}$. We choose a group $G_t$ uniformly at random, and asks each voter to rank all alternatives in $G_t$ according to their stand-alone values. That is, in the final ranking $\rho_i$  from voter $i\in[n]$, for any two alternatives $a\in G_t$ and $b\in G_t$, $\rho_i(a) \leq \rho_i(b)$ if and only if $f_i(\{a\}) \geq f_i(\{b\})$.

We next recall some important notations from the previous section. For an input profile $\hat{\rho}$, the \emph{harmonic score} of $a$ in $\hat{\rho}$ is given by $\textsf{sc}(a, \hat{\rho}) = \sum_{i=1}^n 1/\rho_i(a)$. Let $\hat{f}$ denote the \emph{true} utility profile that is consistent with $\hat{\rho}$ and let $S^*=\arg\max_{S\in \mathcal{F}} \textsf{sw}(S, \hat{f})$ denote the optimal set of alternatives that maximizes the social welfare.

Next, we describe our aggregation rule, which randomly picks a rule from the following two candidates. This, again, aligns with the one introduced in \citep{benade2021preference}.
\begin{itemize}
\item Rule A: Let $G^+_t$ denotes the top (up to) $\sqrt{m}\cdot(1/u_t)$ alternatives from $G_t$ that have the largest harmonic scores in $\hat{\rho}$. Select a subset of $G^+_t$ with a size of $1/u_t$, also chosen uniformly at random.
\item Rule B: Choose a single alternative uniformly at random.
\end{itemize}

Let $G^-_t=G_t\setminus G^+_t$. Note that because $\sqrt{m}\cdot(1/u_0)=\sqrt{m}\cdot m \geq m$, we have $G^+_0=G_0$. Define $G^+=\cup_{t=0}^{\log m} G^+_t$ and $G^-=A\setminus G^+$

Following the same proof of Lemma \ref{lem:y}, we can derive the following lemma.
\begin{lemma}
\label{lem:y1}
The expected social welfare achieved by Rule A is at least  $\textsf{sw}(G^+ \cap S^*, \hat{f})/(1+\log m)\sqrt{m}$.
\end{lemma}

We next show that the expected social welfare achieved under Rule B is at least $(1-c)\textsf{sw}(G^- \cap S^*, \hat{f})/2(1+\log m)\sqrt{m}$.
\begin{lemma}
\label{lem:x1}
Assume the  curvature of each function $f_i$ is at least $c$, the expected social welfare achieved by Rule B is at least $(1-c)\textsf{sw}(G^- \cap S^*, \hat{f})/2(1+\log m)\sqrt{m}$.
\end{lemma}
\emph{Proof:} Observe that $G^-_0=\emptyset$. Therefore,
\begin{eqnarray}
\label{eq:17a}
\textsf{sw}(G^- \cap S^*, \hat{f}) = \textsf{sw}(\cup_{t\in[\log m]}(G^-_t \cap S^*), \hat{f}).
\end{eqnarray}

Fix $t\in [\log m]$ and $a\in G^-_t$. Let $H_k$ be the $k$-th harmonic number, it is easy to verify that
\begin{eqnarray}
\sum_{b\in G_t} \textsf{sc}(b, \hat{\rho}) = n\cdot H_{|G_t|} \leq n\cdot H_{m}.
\end{eqnarray}
Recall that $G^+_t$ contains  the top  $\sqrt{m}\cdot(1/u_t)$ alternatives from $G_t$ that have the largest harmonic scores in $\hat{\rho}$, we have
\begin{eqnarray}
\label{eq:09a}
 \textsf{sc}(a, \hat{\rho})  \leq \frac{n\cdot H_{m}}{\sqrt{m}\cdot(1/u_t)} < \frac{n\cdot (1+\log m)}{\sqrt{m}\cdot m/2^t}.
\end{eqnarray}

Meanwhile, because the  curvature of each function $f_i$ is at least $c$ and $f_i(G_t)\leq 1$ for all each voter $i\in[n]$, we have $\sum_{b\in G_t} f_i(\{b\}) \leq f_i(G_t)/(1-c)  \leq 1/(1-c)$. This, together with the observation that for each alternative $b\in G_t$, $ f_i(\{b\})$ is no larger than the stand-alone value any of its predecessors in $\rho_i$ (this is by the design of Ranking-by-Values), implies that for each voter $i\in[n]$ and alternative $b\in G_t$, we have
\begin{eqnarray}
\label{eq:1a}
f_i(\{b\}) \leq 1/(1-c)\cdot \rho_i(b).
\end{eqnarray}

It follows that for each alternative  $a\in G^-_t$,
\begin{eqnarray}
\label{eq:12a}
\sum_{i\in[n]}f_i(\{a\}) \leq \sum_{i\in[n]} \frac{1}{(1-c)\cdot \rho_i(a)}= \frac{\textsf{sc}(a, \hat{\rho})}{1-c} < \frac{n\cdot (1+\log m)}{(1-c)\cdot \sqrt{m}\cdot m/2^t}
\end{eqnarray}
where the first inequality is by inequality (\ref{eq:1a}) and the second inequality is by inequality (\ref{eq:09a}). By the definition of $G_t$, we have that each alternative  $a\in G^-_t$ has a  cost of at least  $l_t=2^{t-1}/m$. Therefore, for any $t\in[\log m]$, the ``value-to-cost'' density of each alternative $a\in G^-_t$ is at most
\begin{eqnarray}
&&\frac{\textsf{sw}(a, \hat{f})}{l_t} = \frac{\sum_{i\in[n]} f_i(\{a\})}{2^{t-1}/m}\leq   \frac{\frac{n\cdot (1+\log m)}{(1-c)\sqrt{m}\cdot m/2^t}}{2^{t-1}/m} = \frac{2n\cdot (1+\log m)}{(1-c)\sqrt{m}}
\end{eqnarray}
where the inequality is by inequality (\ref{eq:12a}). The rest of the proof is the same as the proof of Lemma \ref{lem:x}, thus omitted here.  $\Box$

Now we are in position to present the main theorem of this section. Given Lemma \ref{lem:y1} and Lemma \ref{lem:x1}, we follow the same proof of Theorem \ref{thm:1} to derive the following theorem.
\begin{theorem}
Assume the  curvature of each function $f_i$ is at least $c$, the expected social welfare achieved by our aggregation rule is at least $(1-c)\textsf{sw}(S^*, \hat{f})/4(1+\log m)\sqrt{m}$.  This also implies that the distortion of our approach is at most $O(\log m\sqrt{m}/(1-c))$.
\end{theorem}

\paragraph{Remark} While we limit each voter to rank alternatives from a random group, the traditional method, which requires each voter to rank \emph{all} alternatives, achieves at least the same distortion. This is because ranking all alternatives offers more information than ranking only a subset. However, this statement does not apply to ranking-by-marginal-values method.

\subsection{Threshold Approval Votes}
Recall that the \emph{threshold approval votes} method asks each voter to report all alternatives whose stand-alone utility is above a certain threshold $\alpha$.  Formally, the approval set for voter $i$ is given by:
\[ \rho_i = \{a \in A \mid f_i(\{a\}) \ge \alpha\}. \]

For $t\in[\log m]$ (for ease of presentation, assume $m$ is a power of $2$), define $l_t=2^{t-1}/m$ and $u_t=2^t/m$. Let $I_0=[0, 1/m]$ and $I_t=(l_t, u_t]$ for $t\in[\log m]$.

Let $\hat{f}$ denote the true utility profile that is consistent with $\hat{\rho}$ and let $S^*=\arg\max_{S\in \mathcal{F}} \textsf{sw}(S, \hat{f})$ denote the optimal set of alternatives that maximizes the social welfare. For $a\in A$ and $t\in \{0, 1, \cdots, \log m\}$, define $d_a^t=|\{i\in [n]\mid  f_i(\{a\})\in I_t\}|$ as the number of voters whose utility of $a$ belongs to the interval $I_t$. We draw inspiration from \citep{benade2021preference} and subsequently present two bounds concerning the social welfare of each alternative $a \in A$:
\begin{eqnarray}
\sum_{t=0}^{\log m} d_a^t \cdot l_t \leq \textsf{sw}(a, \hat{f}) \leq \sum_{t=0}^{\log m} d_a^t \cdot u_t.
\end{eqnarray}

Moreover, because $I_0=[0, 1/m]$, the above bound indicates that\begin{eqnarray}
\label{eq:pp}
\sum_{t=0}^{\log m} d_a^t \cdot l_t \leq \textsf{sw}(a, \hat{f}) \leq \frac{d_a^0}{m}+\sum_{t=1}^{\log m} d_a^t \cdot u_t.
\end{eqnarray}

Now we are ready to introduce our aggregation rule. Our rule consists of two candidate rules:
\begin{itemize}
\item Rule A: The first rule is to compute a set of alternatives through solving a knapsack problem. First, randomly pick a threshold  $\alpha$ from $\{l_1, l_2, \cdots, l_{\log m}\}$ at uniform. Then  assign a weight $w_a$ to each alternative $a\in A$ such that $w_a= |\{i\in [n]\mid f_i(\{a\})\geq \alpha\}|$ represents the number of voters who approve alternative $a$ with respect to the threshold $\alpha$. Then solve the following knapsack problem, which allows a Fully Polynomial Time Approximation Scheme (FPTAS) \citep{vazirani2001approximation}:
     \begin{center}
\framebox[0.6\textwidth][c]{
\enspace
\begin{minipage}[t]{0.6\textwidth}
\small
$\textbf{P.1}$
$\max_{S\subseteq A} \sum_{a\in S} w_a$  subject to $c(S)\leq 1$.
\end{minipage}
}
\end{center}
\vspace{0.1in}
This near-optimal solution is provided as the output from Rule A.
\item Rule B: Choose a single alternative from $A$ randomly at uniform.
\end{itemize}

Recall that $\textsf{sw}(a, \hat{f}) \leq \frac{d_a^0}{m}+\sum_{t=1}^{\log m} d_a^t \cdot u_t$ (by the upper bound in (\ref{eq:pp})), we have
\begin{eqnarray}
\label{eq:kk}
\textsf{sw}(S^*, \hat{f})\leq \sum_{a\in S^*}\textsf{sw}(a, \hat{f}) \leq  \sum_{a\in S^*} (\frac{d_a^0}{m}+\sum_{t=1}^{\log m} d_a^t \cdot u_t) \leq \frac{n}{m}+ \sum_{t=1}^{\log m} \sum_{a\in S^*}  d_a^t \cdot u_t
\end{eqnarray}
where the first inequality is by the assumption that $\textsf{sw}(\cdot, \hat{f})$ is  a submodular function, and the last inequality is by the observation that $\sum_{a\in S^*} d_a^0 \leq n$. In the following two lemmas, we will prove that the expected social welfare of Rule A is close to $\sum_{t=1}^{\log m} \sum_{a \in S^*} d_a^t \cdot u_t$ and the expected social welfare of Rule B is close to $n/m$, respectively. Consequently, randomly selecting a rule from the above two candidates yields a bounded distortion.

\begin{lemma}
\label{lem:a}
Assume the  curvature of each function $f_i$ is at least $c$, the expected social welfare of Rule A is at least $\frac{(1-\epsilon)(1-c)}{2\log m}\sum_{t\in [\log m]} \sum_{a\in S^*}  d_a^{t}\cdot u_{t}$. Here $\epsilon\in[0,1]$ is a number that is arbitrarily close to zero.
\end{lemma}
\emph{Proof:} Recall that in Rule A, we solve the following problem (approximately), here $w_a= |\{i\in [n]\mid f_i(\{a\})\geq \alpha\}|$ and $\alpha$ is a random number from $\{l_1, l_2, \cdots, l_{\log m}\}$:
     \begin{center}
\framebox[0.6\textwidth][c]{
\enspace
\begin{minipage}[t]{0.6\textwidth}
\small
$\textbf{P.1}$
$\max_{S\subseteq A} \sum_{a\in S} w_a$  subject to $c(S)\leq 1$.
\end{minipage}
}
\end{center}
\vspace{0.1in}

  Fix $\alpha=  l_{t}$, we have $w_a= |\{i\in [n]\mid f_i(\{a\})\geq l_{t}\}|$. Now consider a different knapsack problem.
     \begin{center}
\framebox[0.6\textwidth][c]{
\enspace
\begin{minipage}[t]{0.6\textwidth}
\small
$\textbf{P.2}$
$\max_{S\subseteq A} \sum_{a\in S}  d_a^{t}$  subject to $c(S)\leq 1$.
\end{minipage}
}
\end{center}
\vspace{0.1in}

Let $O_2(\alpha)$ denote an optimal solution of $\textbf{P.2}$. By the observation that $S^*$ is a feasible solution of  $\textbf{P.2}$ and $O_2(\alpha)$ denotes an optimal solution of  $\textbf{P.2}$, we have $\sum_{a\in S^*}  d_a^{t} \leq \sum_{a\in O_2(\alpha)}  d_a^{t} $. This indicates that
\begin{eqnarray}
\label{eq:0099}
\sum_{a\in S^*}  d_a^{t} \cdot u_{t} \leq \sum_{a\in O_2(\alpha)}  d_a^{t} \cdot u_{t}.
\end{eqnarray}

  Because $d_a^{t}=|\{i\in [n]\mid  f_i(\{a\})\in (l_{t}, u_{t}]\}|$ and $w_a= |\{i\in [n]\mid f_i(\{a\})\geq l_{t}\}|$, we have  $w_a \geq  d_a^{t}$ for all $t\in A$. Therefore, the value of the optimal solution of $\textbf{P.1}$ is no less than the value of the optimal solution of $\textbf{P.2}$. Let $O_1(\alpha)$ denote the optimal solution of $\textbf{P.1}$, we have
\begin{eqnarray}
\label{eq:9988}
\sum_{a\in O_1(\alpha)} w_a \geq   \sum_{a\in O_2(\alpha)}  d_a^{t}.
\end{eqnarray}

Given that $\textbf{P.1}$ admits a FPTAS, let $V(\alpha)$ denote the solution returned from Rule A, which is a ($1-\epsilon$)-approximation solution of $\textbf{P.1}$. Here $\epsilon\in[0,1]$ is a number that is arbitrarily close to zero. We have:
\begin{eqnarray}
\sum_{a\in V(\alpha)} w_a  \geq (1-\epsilon)\cdot \sum_{a\in O_1(\alpha)} w_a \geq   (1-\epsilon)\cdot \sum_{a\in O_2(\alpha)}  d_a^{t}
\end{eqnarray}
where the second inequality by inequality (\ref{eq:9988}).

This, together with the fact that $l_t\geq u_t/2$ for all $t\in[\log m]$, implies that
\begin{eqnarray}
\sum_{a\in V(\alpha)} w_a \cdot l_{t}  \geq (1-\epsilon)\cdot  \sum_{a\in O_2(\alpha)}  d_a^{t}\cdot \frac{u_{t}}{2}. \label{eq:0000}
\end{eqnarray}

Moreover, it is easy to verify that $\sum_{i\in[n]}f_i(\{a\})\geq  w_a \cdot l_{t}$ for all $a\in A$. Hence,

\begin{eqnarray}
\sum_{a\in V(\alpha)}\sum_{i\in[n]}f_i(\{a\})  \geq \sum_{a\in V(\alpha)} w_a \cdot l_{t}  \geq (1-\epsilon)\cdot  \sum_{a\in O_2(\alpha)}  d_a^{t}\cdot \frac{u_{t}}{2}\geq (1-\epsilon)\cdot  \sum_{a\in S^*}  d_a^{t}\cdot \frac{u_{t}}{2} \label{eq:oo}
\end{eqnarray}
where the second inequality is by inequality (\ref{eq:0000}) and the third inequality is by inequality (\ref{eq:0099}).

Recall that our threshold $\alpha$ is picked randomly from $\{l_1, l_2, \cdots, l_{\log m}\}$ at uniform, we have

\begin{eqnarray}
&&\mathbb{E}_\alpha[\sum_{a\in V(\alpha)}\sum_{i\in[n]}f_i(\{a\})] = \frac{1}{\log m}\sum_{\alpha\in \{l_1, l_2, \cdots, l_{\log m}\}}\sum_{a\in V(\alpha)}\sum_{i\in[n]}f_i(\{a\})   \\
&& \geq \frac{1}{\log m}\sum_{t\in [\log m]} (1-\epsilon)\cdot  \sum_{a\in S^*}  d_a^{t}\cdot \frac{u_{t}}{2}\\
&&=  \frac{1-\epsilon}{2\log m}\sum_{t\in [\log m]} \sum_{a\in S^*}  d_a^{t}\cdot u_{t} \label{eq:00}
\end{eqnarray}
where the inequality is by inequality (\ref{eq:oo}).

Now we are ready to prove this lemma, that is, the expected social welfare of Rule A (i.e., $\mathbb{E}_\alpha[\textsf{sw}(V(\alpha), \hat{f})]$) is at least $\frac{(1-\epsilon)(1-c)}{2\log m}\sum_{t\in [\log m]} \sum_{a\in S^*}  d_a^{t}\cdot u_{t}$. Observe that
\begin{eqnarray}
&&\mathbb{E}_\alpha[\textsf{sw}(V(\alpha), \hat{f})] = \mathbb{E}_\alpha[\sum_{i\in[n]}f_i(V(\alpha))] \\
&&\geq (1-c)\cdot \mathbb{E}_\alpha[\sum_{a\in V(\alpha)}\sum_{i\in[n]}f_i(\{a\})]\\
 &&\geq  \frac{(1-\epsilon)(1-c)}{2\log m}\sum_{t\in [\log m]} \sum_{a\in S^*}  d_a^{t}\cdot u_{t}
\end{eqnarray}
where the first inequality is by the assumption that  the  curvature of each function $f_i$ is at least $c$, and the second inequality is by inequality (\ref{eq:00}). $\Box$

We next analyze the social welfare achieved by Rule B.

\begin{lemma}
\label{lem:b}
The expected social welfare of Rule B is at least $n/m$.
\end{lemma}
\emph{Proof:} Recall that Rule B chooses a single alternative from $A$ randomly at uniform. Therefore, each alternative has a probability of $1/m$ of being selected.  This, together with Lemma \ref{lem:2} and the assumption that $\textsf{sw}(\cdot, \hat{f})$ is a monotone submodular function, implies that
the expected social welfare achieved by  Rule B is at least $\textsf{sw}(A, \hat{f})/m=n/m$. $\Box$

Lemma \ref{lem:a} and Lemma \ref{lem:b} together imply that if we randomly pick a rule from Rule A and Rule B, then the expected social welfare is at least
\begin{eqnarray}
\frac{1}{2} \cdot (\frac{(1-\epsilon)(1-c)}{2\log m}\sum_{t\in [\log m]} \sum_{a\in S^*}  d_a^{t}\cdot u_{t} +  \frac{n}{m}).
\end{eqnarray}

Observe that
\begin{eqnarray}
&&\frac{1}{2} \cdot (\frac{(1-\epsilon)(1-c)}{2\log m}\sum_{t\in [\log m]} \sum_{a\in S^*}  d_a^{t}\cdot u_{t} +  \frac{n}{m})\\
&&\geq \frac{(1-\epsilon)(1-c)}{4\log m} \cdot (\sum_{t\in [\log m]} \sum_{a\in S^*}  d_a^{t}\cdot u_{t} +  \frac{n}{m})\\
&&\geq \frac{(1-\epsilon)(1-c)}{4\log m} \cdot \textsf{sw}(S^*, \hat{f})
\end{eqnarray}
where the second inequality is by inequality (\ref{eq:kk}). This leads to the following main theorem.

\begin{theorem}
Assume the  curvature of each function $f_i$ is at least $c$, the expected social welfare achieved by our aggregation rule is at least $(1-\epsilon)(1-c)\textsf{sw}(S^*, \hat{f})/4\log m $. Here $\epsilon\in[0,1]$ is a number that is arbitrarily close to zero. This also implies that the distortion of our approach is at most $O(\log m/(1-c))$.
\end{theorem}

\paragraph{Remark}Note that in the special case where \(c = 0\), assuming additive utility functions, our distortion bound is reduced to \(O(\log m)\). This improves upon the state-of-the-art result of \(O(\log^2 m)\) presented in \citep{benade2021preference}. This improvement is partly because our algorithm does not require partitioning alternatives based on their costs, as was done in the previous work. On the hardness side, \cite{benade2021preference} constructed an example demonstrating that the distortion associated with the threshold approval voting method is at least \(\Omega(\log m / \log \log m)\).

\section{Conclusion}
We extend existing work by proposing a submodular participatory budgeting problem with monotone submodular utility functions. We evaluate three preference elicitation methods, focusing on distortion. Our aggregation rule for threshold approval votes outperforms the state-of-the-art for additive utility functions.

As previously mentioned, our submodular setting can only capture negative interactions among alternatives. We aim to extend this study to incorporate both negative and positive interactions in the future. Additionally, our current study focuses on a randomized approach; exploring deterministic solutions in the same context would be an interesting direction for future research.

\bibliographystyle{ijocv081}
\bibliography{reference}




\end{document}